\documentclass{emulateapj}
\usepackage{amsmath}
\usepackage{graphicx}
\usepackage{natbib}
\usepackage{bm}

\newcommand{\rc}{\ensuremath{r_{\rm core}}}
\newcommand{\rL}{\ensuremath{r_{\rm L}}}
\newcommand{\rt}{\ensuremath{r_{\rm ta}}}
\newcommand{\rfs}{\ensuremath{r_{\rm fs}}}
\newcommand{\Mfs}{\ensuremath{M_{\rm fs}}}

\begin{document}

\title{Cores and Cusps in Warm Dark Matter Halos}
\author{Francisco Villaescusa-Navarro}
\affiliation{Instituto de Fisica Corpuscular (IFIC), E-46071, Valencia, Spain}
\author{Neal Dalal}
\affiliation{Canadian Institute for Theoretical Astrophysics, Univ.\
  of Toronto, 60 St.\ George St., Toronto, ON, M5S3H8, Canada}

\shorttitle{Cores and Cusps in WDM Halos}
\shortauthors{Villaescusa-Navarro \& Dalal}

\begin{abstract}
The apparent presence of large core radii in Low Surface
Brightness galaxies has been claimed as evidence in favor of warm
dark matter.  Here we show that WDM halos do not have cores that are
large fractions of the halo size: typically, 
$\rc / r_{\rm 200} \lesssim 10^{-3}$.  This suggests an
astrophysical origin for the large cores observed in these galaxies,
as has been argued by other authors.
\end{abstract}

\section{Introduction}

The highly successful cold dark matter (CDM) model idealizes the
thermal motions of dark matter particles as negligible on all scales
at high redshift.  In this model, perturbation modes on all scales are
gravitationally unstable, leading to hierarchical structure formation
in which nonlinear structures such as halos assemble through numerous
mergers.  Numerical simulations of structure formation within CDM
models indicate that halos are predicted to have steep central density
profiles, with logarithmic slopes $d\log\rho/d\log r \sim -1$ on
the smallest resolved scales (see \citet{aquarius} for a recent
example).  

In many DM models, however, the DM temperature is nonzero, which can
affect the properties of DM halos in multiple ways.  For example, a
finite DM temperature suppresses the abundance of low-mass halos.
This occurs because, following freezeout of dark matter
interactions, DM particles freely stream over some distance determined by
their thermal velocities, and density fluctuations on scales below
this free-streaming scale $\rfs$ are highly suppressed.  Roughly
speaking, the smallest halos that arise are expected to have masses of
order $\Mfs=(4\pi/3){\bar\rho_m}\rfs^3$, although N-body
simulations have not definitively ruled out the formation of at least
some halos below $\Mfs$ through non-hierarchical processes like
fragmentation \citep{Wang07}.  

Besides this suppression of the abundance of halos, a nonzero DM
temperature also can affect the central density profiles of the halos
that do form.  One elegant way to see this is to note that the DM
phase space density is finite if the DM temperature is finite, and
since DM is taken to be collisionless, then Liouville's theorem
guarantees that the phase space density cannot increase.  Hence, the
phase space density is bounded within DM halos \citep{Tremaine79},
which implies that the central slope of the DM density profile must
vanish \citep[e.g.][]{Tremaine94}.  In other words, halos are expected
to have central cores if the DM is not cold.  Note that this effect is
caused by the finite DM temperature at the time of formation of the
halo, and is not due to the truncation of the small-scale linear power
spectrum \citep{Wang09}.

Therefore, increasing the DM temperature has the effect of suppressing
the number of low-mass halos, and of producing central cores in DM
halos.  Observationally, there may be evidence for both of these
effects.  The observed number of Local Group satellite galaxies falls
well below the thousands of DM subhalos found in CDM simulations of
halos like our Galaxy's (see \citet{Kravtsov10} for a recent review).
In addition, the 21 cm rotation curves of certain galaxies, in
particular low surface brightness (LSB) galaxies, appear better fit by
cored DM profiles than cuspy DM profiles (see \citet{Kuzio09} for a
recent discussion).  For these reasons, there has been considerable
interest in the literature in investigating structure formation in
models where DM is not perfectly cold.  In so-called Warm Dark Matter
(WDM) models, the DM temperature is chosen to make the free-streaming
scale correspond to subgalactic scales, $\rfs\sim 0.1 h^{-1}{\rm Mpc}$
\citep{Bode01}.

WDM models have become increasingly disfavored in recent years, in
large part because of constraints on the matter power spectrum derived
from the Lyman-$\alpha$ forest flux power spectrum \citep{Seljak06}.
The observational support for WDM models from dwarfs and LSB galaxies
has also eroded, as faint Local Group satellites have been discovered
in increasing numbers
\citep[e.g.][]{Belokurov07,Tollerud08,Kravtsov10} thanks to the Sloan
Digital Sky Survey.  Recently, \citet{Kuzio09} have argued that
the large cores apparently observed in certain LSB galaxies cannot all
be due to WDM, since the implied central phase space densities in
these systems are not universal, but instead show large variations
from object to object.  This suggests an astrophysical origin for
claimed detections of central cores, and mechanisms to produce such
cores have been proposed \citep[e.g.][]{Governato09}.

In this paper, we present yet another argument against WDM as the
origin of large cores in halo density profiles.  As discussed by
\citet{Kuzio09}, the inferred core radii in several LSB galaxies are
large fractions of the halo virial radii, $\rc \sim 5\%\,
r_{\rm 200}$.  As mentioned above, such large cores do not arise in
CDM models, and so we might naturally consider producing large cores
by raising the DM temperature.  Making the DM warm, however, has the
side effect of wiping out small halos, and so it is not obvious that
raising the DM temperature can generate halos with large 
$\rc/r_{\rm 200}$.  

We can, however, use a simple argument
to make an order-of-magnitude estimate of this ratio.
Consider a WDM particle of mass $m$ and typical momentum $p$.
Following freezeout, its momentum redshifts as $p\propto a^{-1}$, so
let us write $p=m v_0 / a$, where $v_0$ is the velocity today at $z=0$
(since WDM must be nonrelativistic today, $v_0\ll 1$).  The particle's
velocity is then $v(a) = v_0/\sqrt{v_0^2+a^2}$ (using units where
$c=1$).  Neglecting accelerations caused by gravitational potential
fluctuations, the particle freely streams a comoving distance
\begin{equation}
\rfs = \int \frac{v\,dt}{a} \sim
\frac{v_0}{\Omega_r^{1/2}H_0} \log\left(\frac{a_{\rm eq}}{v_0}\right),
\end{equation}
where we assume $v_0\ll a_{\rm eq}=\Omega_r/\Omega_m$.  This distance
encloses mass $\Mfs=(4\pi/3){\bar\rho_m}\rfs^3$, and as noted above,
the smallest halos that form will have masses of order $\Mfs$.  The
virial velocities of these halos at their formation epoch $a_c$ are
\begin{equation}
v_{\rm 200}=\left(\frac{G\Mfs}{r_{\rm 200}}\right)^{1/2}=
\left(\frac{\Omega_m\Delta_{\rm 200}^{1/3}}{2a_c}\right)^{1/2}H_0\,\rfs,
\end{equation}
where $\Delta_{\rm 200}\approx 200$ is the virial overdensity.
Now, because the thermal velocity $v_{\rm th}\approx v_0/a_c$ at
expansion factor $a_c$ is nonzero, infalling particles will not fall
directly towards the halo center, but instead have a nonzero impact
parameter, and the typical impact parameter determines the core size.
Naively, we might expect that at formation ($a=a_c$),
\begin{equation}
\frac{\rc}{r_{\rm 200}} \sim \frac{v_{\rm th}}{v_{\rm 200}}\sim
\left(\frac{\Omega_r}{\Omega_m}\frac{2}{\Delta_{\rm 200}^{1/3}a_c}
\right)^{1/2} \!\!\! /\log\left(\frac{a_{\rm eq}}{v_0}\right).
\label{equa}
\end{equation}
Following formation, any subsequent growth in halo mass can only
decrease the core radius, while the virial radius can only increase.
Indeed, even if the mass distribution around the halo is static, with
no accretion following formation, the virial radius will still grow in
time proportionally to the expansion factor $a$, because the expansion
of the universe dilutes the background mean matter density.  This is
the reason why halo concentrations correlate with halo formation times
in CDM cosmologies \citep[e.g.][]{Wechsler02,Zhao03a,Zhao03b}. 
Therefore, following the formation epoch $a_c$, the ratio 
$\rc$/$r_{\rm vir}$ must diminish in time at least as fast as
$a_c/a$; any mass accretion will only decrease this ratio even faster.
Assuming no growth, then at the present time ($a=1$) 
\begin{equation}
\frac{\rc}{r_{\rm 200}} \sim
\left(\frac{\Omega_r}{\Omega_m}\right)^{1/2}
\left(\frac{2a_c}{\Delta_{\rm 200}^{1/3}}\right)^{1/2}/
\log\left(\frac{a_{\rm eq}}{v_0}\right).
\end{equation}
This ratio is maximized by delaying halo formation as late as
possible.  For typical parameters, and setting $a_c=1$, this gives
$\rc \approx 10^{-3} r_{\rm 200}$ observed today.

From this simple order-of-magnitude estimate, it appears unlikely that
WDM models can produce sufficiently large core radii to explain LSB
galaxies.  This argument is only approximate, however.  To make
further progress, we have performed calculations of halo formation in
WDM models.  Our results indicate that WDM halo cores are broadly
consistent with (though typically smaller than) the above estimate,
which precludes WDM as an explanation for the large cores that are
claimed to exist in certain LSB galaxies.

\section{Numerical Method}

In this section we describe our numerical method to solve for the
self-consistent halo density profile following collapse.  We eschew
conventional N-body simulations, since for feasible simulation
parameters the core radii will typically be unresolved or (at best)
marginally resolved \citep{Colin08}.  Since we are interested in
studying the behavior on scales smaller than the typical resolution
limits of conventional N-body simulations, we have instead
employed an alternative approach similar to that used by \citet{LD10}.

We calculate the collapse of isolated peaks in an expanding
universe.  To isolate the effects of the nonzero WDM temperature
during halo collapse, we focus on the case of spherical collapse.
This problem has been investigated previously in the literature, and
it is straightforward to show that cold, spherical collapse gives 
halos with central density profiles behaving as $\rho\propto r^{-2}$,
or steeper \citep{Fillmore84,Bert85,LD10}.  Because cold spherical
collapse is well-understood, any departures from $r^{-2}$ profiles are
clearly due to the effects of warm collapse.  By Newton's theorem,
our calculations neglect the effects of the local environment of
peaks, which cause peaks to collapse nonspherically.  We know from
previous cosmological WDM simulations, however, that accounting for
the effects of local environment does not lead to large cores in WDM
halos \citep{Colin08,Wang09}.  

For a given potential $\Phi(r,t)$, assumed to be spherically
symmetric, we integrate the equations of motion to solve for the orbit
$R(t)$ for each particle. Given the orbit $R(t)$, we compute the mass
profile deposited by each particle. We compute orbits for many
particles, and sum over all their deposited mass profiles to obtain
the total mass profile $M_{\rm total}(r,t)$, and the total density
$\rho=(dM_{\rm total}/dr)/(4\pi r^2)$.  Then, we repeat this
procedure, using the newly obtained mass profile, and iterate to
convergence.

We initialize this procedure using linear perturbation theory.  We
start with a linear density profile $\delta(\rL)$ describing the
initial peak at the starting epoch $a_{\rm init}=(1+z_{\rm init})^{-1}$.  
Here, \rL\ is a comoving Lagrangian radius, to be distinguished
from the proper Eulerian radius $r$ at subsequent times.
We choose the initial peak profile to be proportional
to the (linear theory) matter correlation function, 
$\delta(\rL)\propto\xi(\rL)$.  This
corresponds to the average profile of high peaks in the
$\nu\to\infty$ limit \citep{BBKS}, and so this profile should be
typical of the first halos to form in WDM cosmologies.  The matter
correlation function depends on the WDM transfer function, which
\citet{Bode01} found to be well described by the parametrization
\begin{equation}
T_{\rm WDM}(k) = [1+(\alpha k)^{2\nu}]^{-5/\nu},
\end{equation}
where $\nu\approx 1.2$ and $\alpha$ is a characteristic length scale.
We parametrize WDM models by their free-streaming scale \rfs, or
equivalently the enclosed mass \Mfs, so we require a translation
between \rfs\ and $\alpha$.  We determine the equivalent
free-streaming length for a given $\alpha$ by matching the
\citeauthor{Bode01} transfer function to top-hat smoothing, which is
given by
\begin{equation}
W_{\rm TH}(k)=\frac{3}{(k R)^3}[\sin(k R) - k R \cos(k R)]
\end{equation}
for smoothing scale $R$, which we take to be equal to \rfs.  We have
found that $\alpha\approx 5\rfs$ provides a reasonable match to the
two functions.  

Given a desired free-streaming mass \Mfs\ and halo formation redshift
$z_{\rm form}$, we set the initial peak profile shape 
$\delta(r=a_{\rm init}\,\rL)$ to be
proportional to the correlation function (using the appropriate \rfs),
and normalize the peak height at the starting redshift so that 
the average interior overdensity
${\bar\delta}\equiv 3 r^{-3}\int_0^r r^2 \delta\,dr$,
linearly evolved to redshift $z_{\rm form}$ and evaluated at the
free-streaming scale is
${\bar\delta}(\rL=\rfs, z=z_{\rm form})=\delta_c=1.686$,
in accordance with the spherical collapse model \citep{GunnGott72}.
We assume that this linear profile evolves at early
times according to linear perturbation theory: 
\begin{equation}
\delta(r=a\,\rL,a)=D(a)\delta(\rL)
\end{equation}
where $D(a)$ is the linear growth factor, which for $\Lambda$CDM
universes with no pressure perturbations may be expressed as
\begin{equation}
D(a)\propto H(a)\int_0^{a}\frac{da}{(aH)^3},
\end{equation}
normalized so that $D(a=1)=1$ \citep{Peebles80}.  Note that 
this procedure is not entirely self-consistent, since our use of CDM
growth factors neglects the scale-dependence in WDM growth factors
caused by residual free-streaming at late times.  Our neglect of this
residual free streaming when normalizing the initial peak height means
that our peaks do not actually reach $\delta=\delta_c$ at redshift
$z_{\rm form}$, leading to slight errors in the formation epoch.  This
does not appear to affect our results significantly.

This procedure specifies the initial overdensity perturbation at the
starting redshift of the simulation.  We also require the initial
velocities for all the particles.  These velocities have three
contributions: the Hubble velocity, the bulk peculiar velocity, and a
random thermal velocity.  The Hubble term is of course just given by
$\bm{v}_H=H\,\bm{r}$.  The bulk peculiar velocity may be computed from
the density profile, using the linearized continuity equation
\begin{equation}
\dot{\delta}+\bm{\nabla}\cdot\bm{v}=0,
\end{equation}
along with the assumption of potential flow at early times
(i.e.\ $\nabla\times\bm{v}=0$).  This gives
\begin{equation}
v_r(r,a)\simeq -\frac{1}{3}r\bar{\delta}(r,a)H(a),
\end{equation}
where again ${\bar\delta}\equiv 3 r^{-3}\int_0^r r^2 \delta\,dr$.
In addition to this bulk peculiar velocity, for each particle we add a
thermal velocity, drawn from a Fermi-Dirac distribution function for
WDM temperature $T$:
\begin{equation}
f(p,T)d^3p=\frac{1}{N_0(T)}\frac{p^2dp}{e^{p/k_BT}+1}
\label{df}
\end{equation}
where the normalization is given by
\begin{equation}
N_0(T)=\int_0^\infty\frac{p^2dp}{e^{p/T}+1}.
\end{equation}
We typically begin at redshift $z=100$.  
We sample 6000 initial radii spaced
uniformly in volume, up to a maximum radius chosen to enclose the
virialized region at $z=0$.  For each radius, we sample the momentum
distribution with 500 points and the angular distribution with 200
points, weighting the particles according to the fraction of initial
volume, solid angle, and momentum distribution that they represent.

Given the initial conditions for each particle, we then integrate
forward the equations of motion using a fourth order, variable
timestep Runge-Kutta integrator.  The equations of motion are given by
the usual Newtonian dynamics:
\begin{equation}
\frac{d^2R}{dt^2} - \frac{L^2}{R^3} = -\nabla\Phi,
\label{neqm}
\end{equation}
where the angular momentum $\bm{L}=\bm{R}\times\bm{v}$ is conserved
because of the assumed spherical symmetry.  


Given an orbit $R(t)$, the enclosed mass profile deposited by each
particle is  
\begin{equation}
M(r,t) = m_p \Theta[r - R(t)],
\label{deposition}
\end{equation}
where $m_p$ is the mass represented by the particle, and $\Theta(x)$
is the step function.  Summing over all particles gives the total mass
profile $M_{\rm total}(r,t)$, and the density $\rho(r,t)$ and potential
$\Phi(r,t)$ follow easily.  Having computed the mass profile $M(r,t)$
for a given iteration, we then use that mass profile in the equations
of motion for the subsequent iteration.  In practice, we bin the mass
profile using a grid with 350 bins spaced uniformly in expansion
factor $a$ and 500 logarithmically spaced bins in radius,
and then linearly interpolate from this grid to estimate the mass $M$
at arbitrary times and radii as needed for the orbit integrations.

\begin{figure}
\centerline{\includegraphics[width=0.47\textwidth]{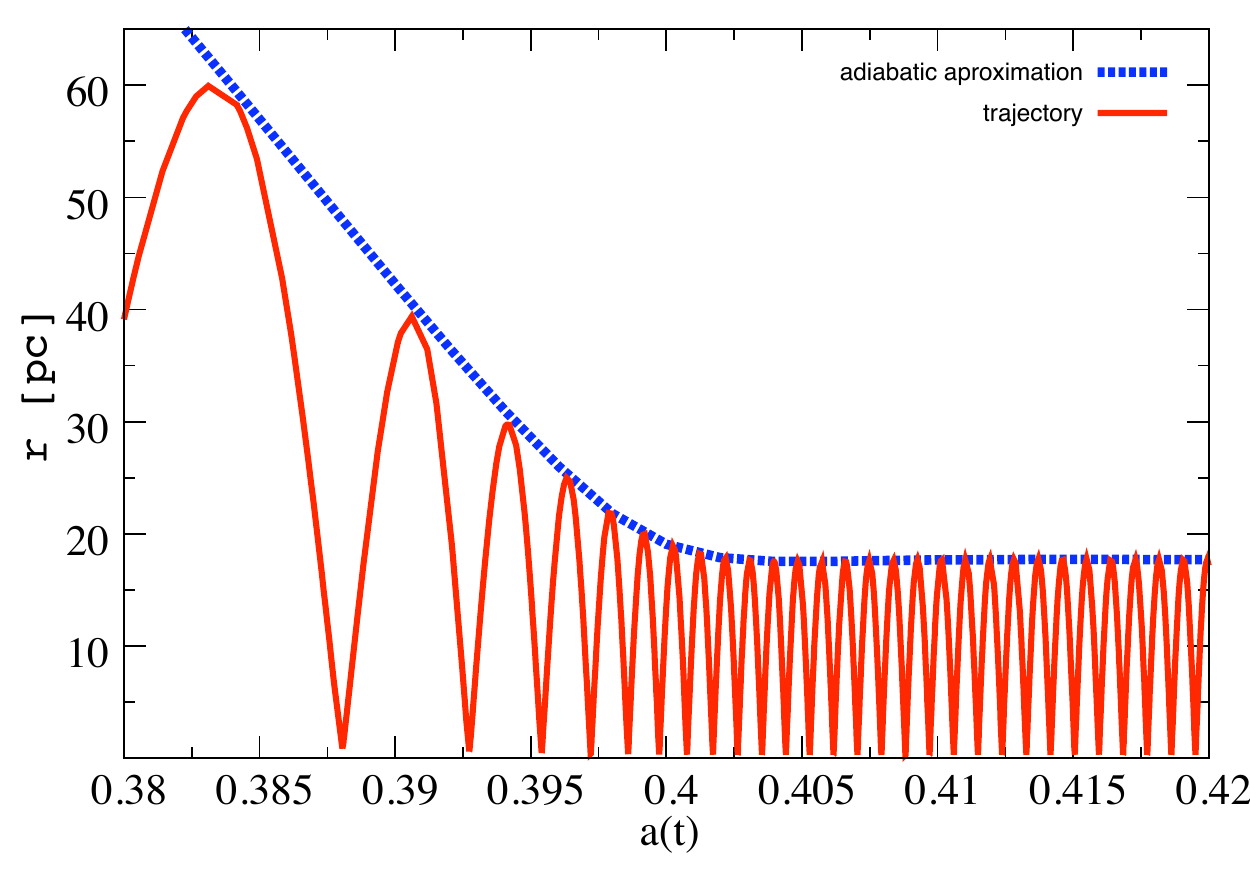}}
\caption{Adiabatic orbital evolution.  The solid red curve shows an example
  orbit $r(t)$ for a particle from one of our collapse calculations.
  For comparison, the dashed blue curve shows the expected behavior for the
  apoapse under the assumption that the orbit responds adiabatically
  to the deepening gravitational potential. 
\label{fig:adiabatic}
}
\end{figure}

To expedite this calculation, we have made use of a simplifying
approximation.  For particles deep within the halo, whose orbital
times are small compared to the Hubble time, we stop the orbital
integration after the fractional change in the product
$R_{\rm apo}\times M(R_{\rm apo})$ over one orbit is less than
$10^{-3}$. Thereafter, we assume that the orbit evolves
adiabatically. Specifically, we assume that the radial action 
$J_r=\oint v_r dR \propto [R_{\rm apo} M(R_{\rm apo})]^{1/2}$ 
is an adiabatic invariant.  Given the time evolution of the mass
profile $M(r,t)$, we then easily determine the time evolution of the
orbital apoapse $R_{\rm apo}$.  As Fig.\ \ref{fig:adiabatic}
illustrates, adiabaticity is an excellent approximation for these
orbits.  Similarly, we assume that the ratio of
periapse to apoapse, $R_{\rm peri}/R_{\rm apo}$, is also conserved
because of conservation of angular momentum.  Given $R_{\rm peri}$ and
$R_{\rm apo}$, we then assume that the mass profile deposited by this
particle is 
\begin{equation}
M(r,t)=m_p\times
\begin{cases}
1, & \mbox{if } r>R_{\rm apo} \\
\frac{r-R_{\rm peri}}{R_{\rm apo}-R_{\rm peri}}, & 
\mbox{if } R_{\rm peri}<r<R_{\rm apo}\\
0, & \mbox{if } r<R_{\rm peri} \\
\end{cases}
\end{equation}
which is a good approximation to Eqn.~(\ref{deposition}), except for
radii very near $R_{\rm peri}$ or $R_{\rm apo}$.

\begin{figure}
\centerline{\includegraphics[width=0.47\textwidth]{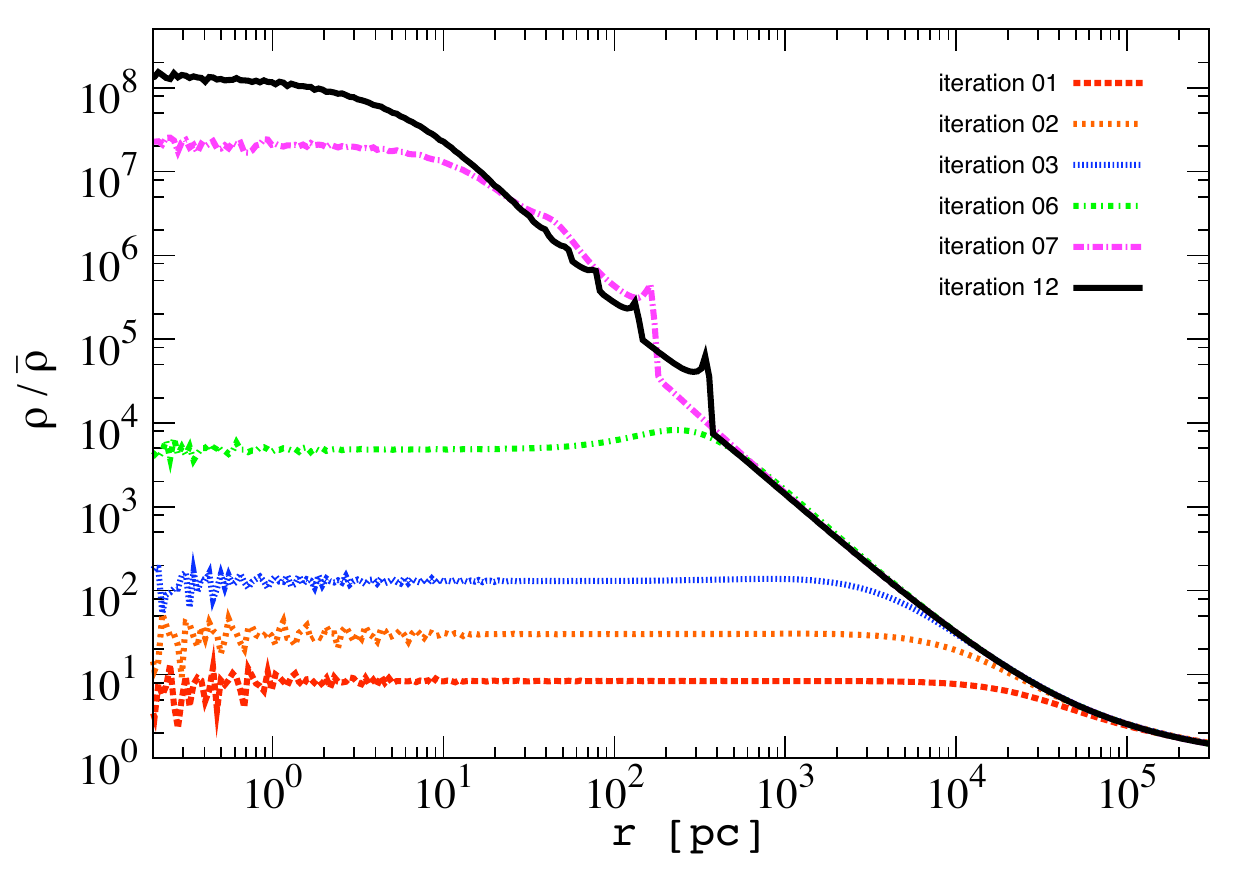}}
\caption{Convergence of the density profile.  The halo density profile
  at $z=2.7$ after iterations 1,2,3,6 and 7 are shown in the colored curves,
  while the black curve shows the converged profile (12 iterations).
\label{convergence}}
\end{figure}

This iterative procedure rapidly converges to a self-consistent
collapse solution; Figure \ref{convergence} illustrates one typical
example.  As the figure shows, the interior density profile quickly
settles into roughly $r^{-2}$ behavior, as expected, although there
are features at both large radii and small radii.  The spikes at large
radii are caustics, a well-known feature of cold spherical collapse
\citep{Fillmore84,Vogelsberger09}.  At small radii, discreteness
effects of the finite number of particles leads to noise in the
determined profile.  This noise in the mass profile enters the
equations of motion for particles, which affects particle orbits and
leads to rapidly developing instabilities in the phase-space structure
\citep{Henon73,Barnes86,Henriksen97}.  These instabilities
significantly distort the shape of the radial caustics at times
following collapse.  We have checked that if we suppress these
instabilities by artificially smoothing the potential, the caustics
match the expected form.  Because these instabilities are physical,
rather than numerical in origin, we have opted not to suppress them.
Accordingly, our density profiles at late times, long after collapse,
do not show the expected prominent caustics.

\section{Results}

In this section we present results of our calculations.  In the first
subsection, we illustrate the behavior found in one typical
simulation, and in the following subsection we describe how the
behavior changes as we vary several physical parameters.  

\subsection{Anatomy of a WDM halo}

\begin{figure}
\centerline{\includegraphics[width=0.47\textwidth]{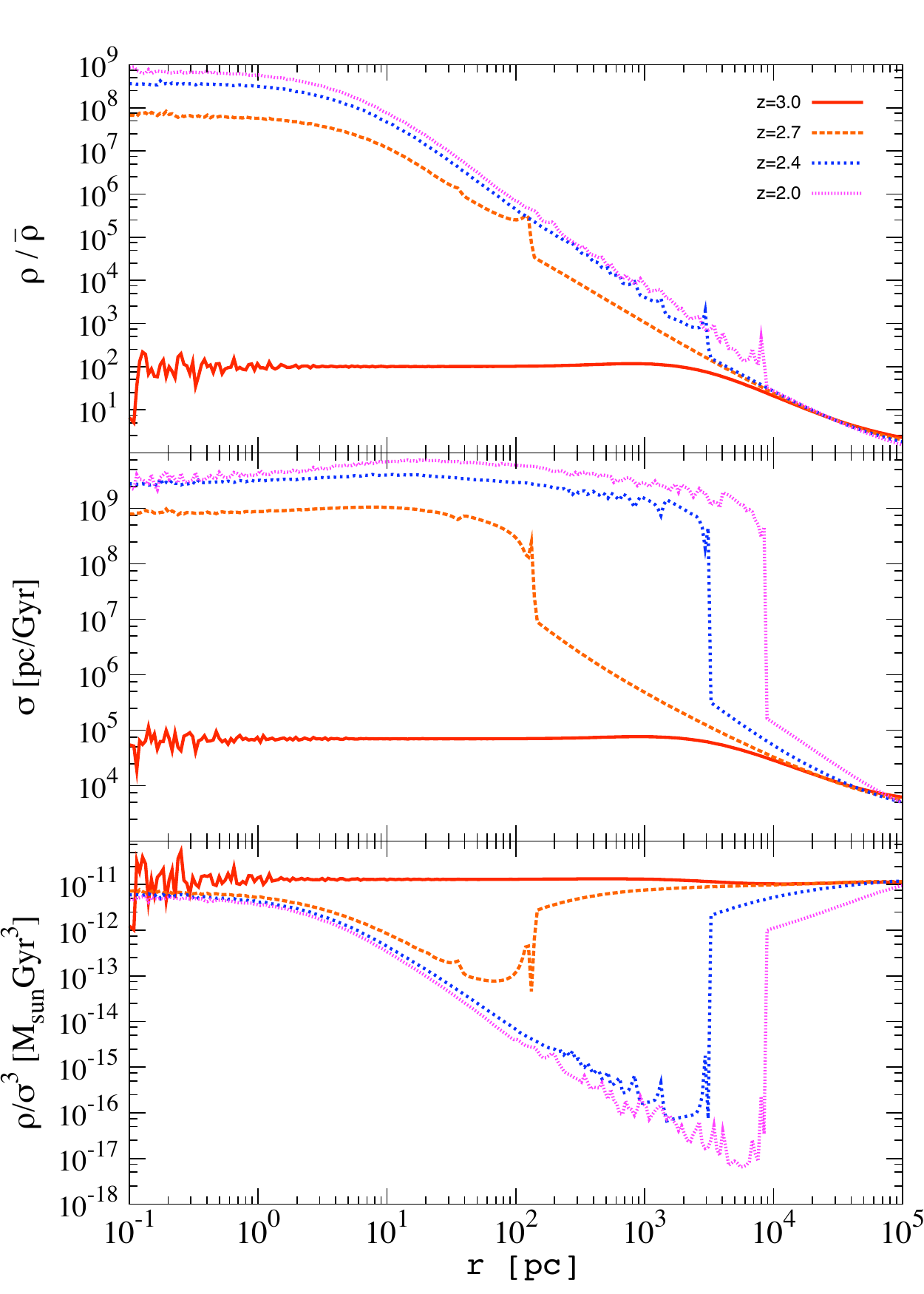}}
\caption{Collapsed profile as a function of time.  The top panel shows
  the halo density profile $\rho(r)$, in units of the mean matter
  density ${\bar\rho}$, at various redshifts before, during, and after
  collapse at $z\sim 2.7$.  The middle panel shows the 3-D velocity
  dispersion.  The bottom panel shows the pseudo-phase-space density
  $\rho/\sigma^3$.  
\label{phasespace}}
\end{figure}

\begin{figure}[t]
\centerline{\includegraphics[width=0.47\textwidth]{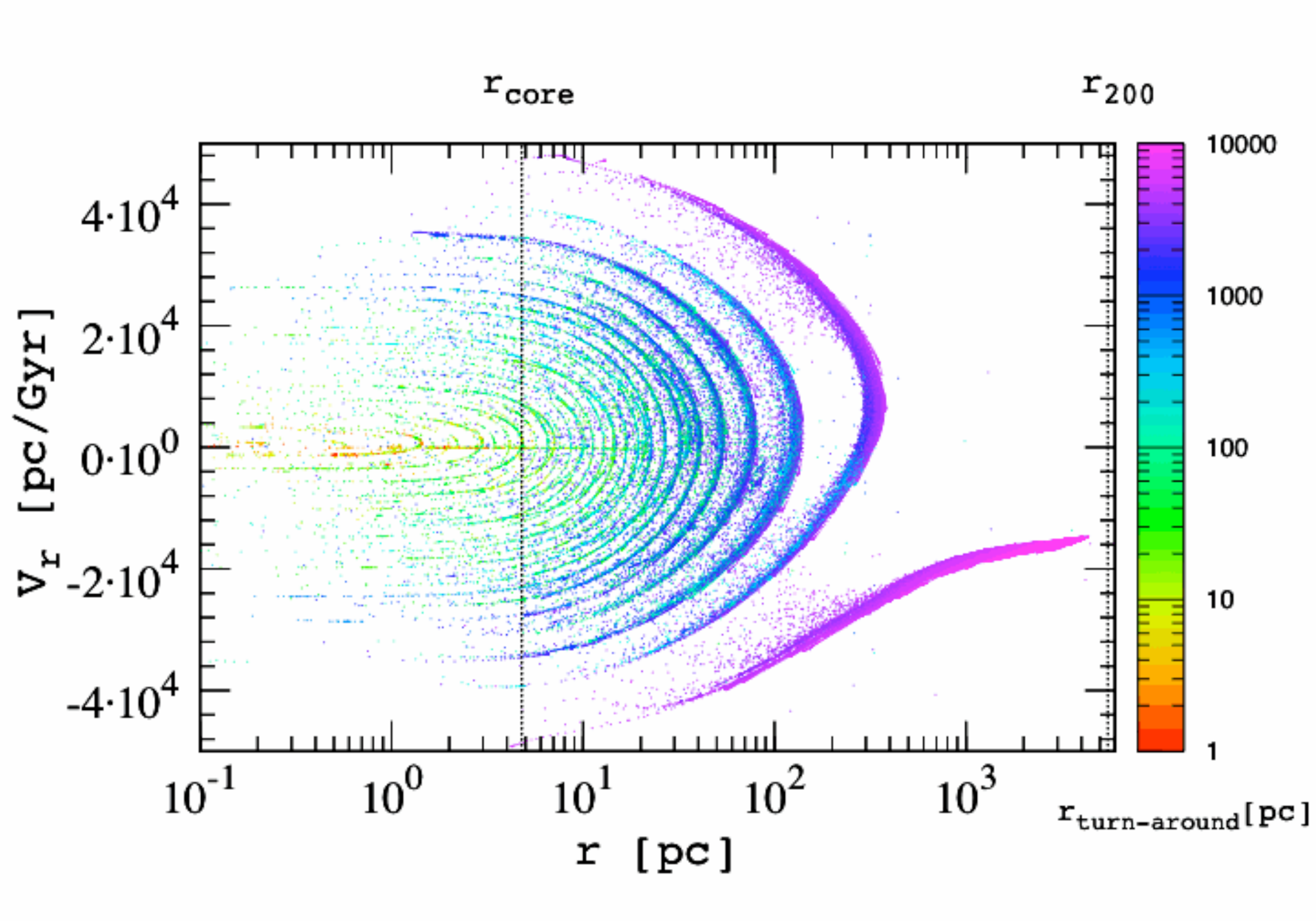}}
\caption{Radial behavior.  Depicted are the radii $r$ and radial
  velocities $v_r$ for a subset of particles at $z=2.7$.  The
  color for each point indicates each particle's turnaround radius.
  The vertical dotted lines indicate \rc\ and $r_{200}$ at this redshift.
\label{streams}}
\end{figure}

In this subsection, we describe the detailed results of one of our
simulations.  The behavior found for this halo is representative of
our simulations.  For concreteness, we use a WDM temperature
$T_0/m=1.3\cdot10^{-8}$ in units where $c=1$, which gives a
free-streaming scale $\rfs\sim 100\, h^{-1}$ kpc, and normalize the
initial peak amplitude so that halo formation occurs near redshift
$z=3$.  Figures \ref{phasespace} and \ref{streams} show the
results.  Figure \ref{phasespace} plots the
density profile $\rho(r)$, the 3-D velocity dispersion $\sigma(r)$,
and the pseudo-phase-space density $\rho/\sigma^3$, as a function of
time, while Fig.\ \ref{streams} illustrates a snapshot in time of the
radial dynamics. At early times, prior to
collapse, the density evolves perturbatively, so that the $\rho(r)$
profile is similar to the original linear density perturbation
$\delta(r)$, simply growing in amplitude. The phase-space density
remains very homogeneous before collapse.  Near the time of collapse,
there is smooth infall towards the halo, and both the density and
velocity dispersion rise in concert to keep $\rho/\sigma^3$ nearly
constant.  Orbits do cross in the infall region, however, since
particles with different thermal velocities fall into the collapsing
halo at different rates.   

The initial collapse of the halo produces a roughly $r^{-2}$ density
profile, due to our assumption of spherical symmetry.  This breaks to
a shallower $\rho\sim\,$const behavior at the 
core radius $\rc$, where the infalling particles reach
periapse.  Following periapse, the particles splash back outwards
with positive radial velocity.  The splash-back surface defines
the outermost caustic, where both the density and velocity dispersion
jump precipitously.  The density jumps at the caustic due to a pile-up
of particles with similar apoapses.  The velocity dispersion jumps
because outside the caustic, particles are all falling inwards,
whereas inside the caustic there is both outwards and inwards motion.
Inside the caustic, the velocity dispersion remains roughly isothermal,
at a value near the halo's virial velocity.  

Because $\rho\sim r^{-2}$ and $\sigma$ is nearly constant with radius,
the pseudo-phase space density shows a nearly power-law behavior over
much of the virialized region.  Outside the outermost caustic,
$\rho/\sigma^3$ remains nearly identical to the phase space density of
the unperturbed material.  At the caustic, $\rho/\sigma^3$ falls
steeply, due to the sudden increase in velocity dispersion.  Towards
smaller radii, $\rho/\sigma^3$ rises smoothly, close to $r^{-2}$.  It
is important to stress, however, that $\rho/\sigma^3$ is not a good
proxy for the actual phase-space density over much of the halo's
interior.  The velocity dispersion tensor is highly anisotropic, in
the sense that radial velocities are much larger than tangential
velocities, as can be seen from the predominantly radial orbits shown
in Figures \ref{fig:adiabatic} and \ref{streams}.  For this reason,
$\sigma\approx\sigma_r$, and so 
$\sigma^3\approx\sigma_r^3\gg\sigma_r\sigma_\theta\sigma_\phi$.  
Only near the core radius does the velocity dispersion become close to
isotropic.  

Many of these features are similar to what is found in cold spherical
collapse \citep{Fillmore84,Bert85}.  The most obvious difference
between warm and cold collapse is
the presence of a core radius, caused by the orbits' inability to
reach $r=0$ due to their nonzero angular momenta.  We estimate the
core radius by fitting the density profile to the functional form
\begin{equation}
\rho=\frac{\rho_c}{\left[1+\left(\frac{r}{\rc}\right)^\alpha\right]^{2/\alpha}}.
\label{fit}
\end{equation}
The parameter $\alpha$ controls how sharply the profile breaks from
$r^{-2}$ behavior to constant density, and typically our simulations
give $\alpha\approx 1-2$.  We are mainly interested, however in the core
radius.  As Figure \ref{streams} shows, $\rc$ coincides with the
typical location of the periapse of the infalling particles.  Both the
density and pseudo-phase space density plateau at $\rc$, the latter
saturating at a value near the phase space density of unvirialized
material outside the halo as expected from the \citet{Tremaine79}
bound.

\begin{figure}
\centerline{\includegraphics[width=0.47\textwidth]{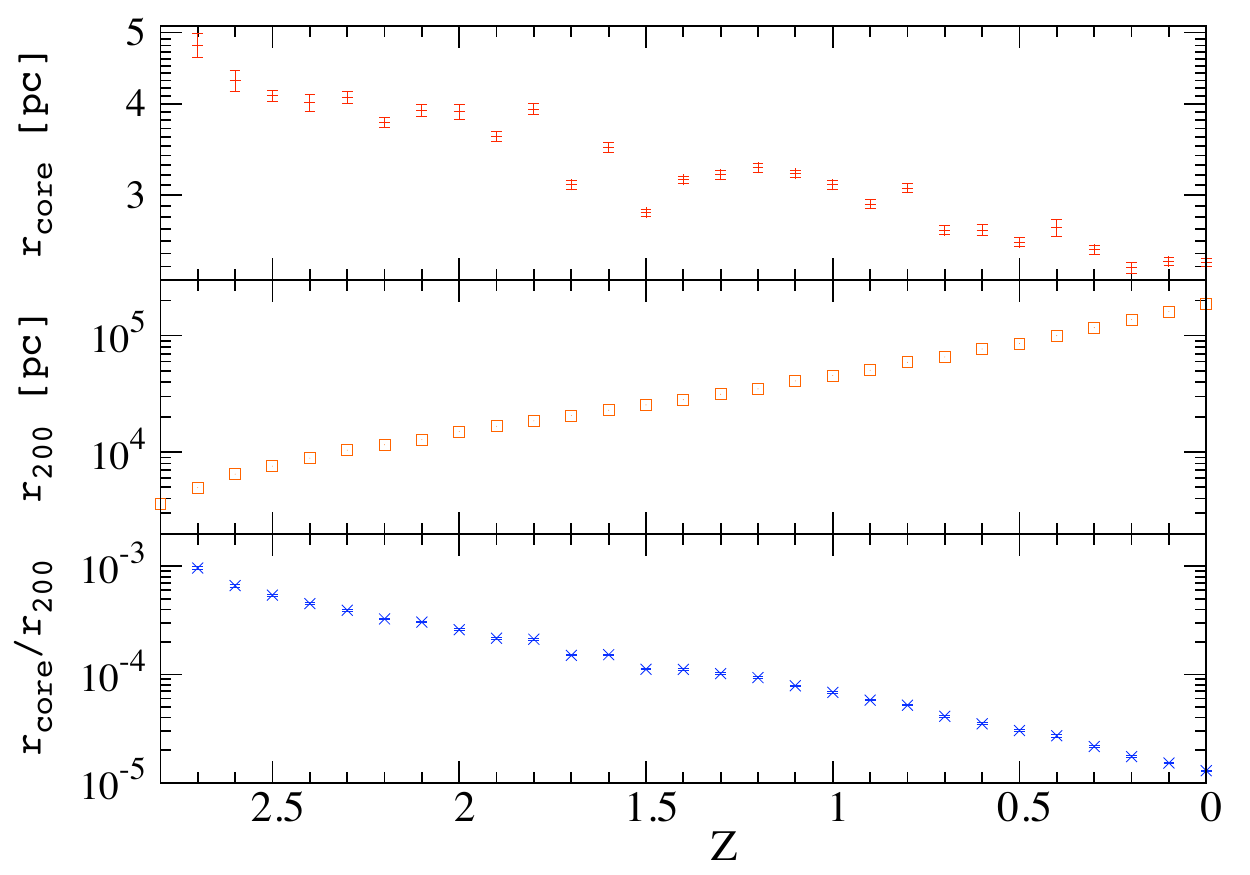}}
\caption{
  Time dependence of halo properties.  The upper, middle and
  lower panels show \rc, $r_{200}$, and the ratio $\rc/r_{200}$
  respectively, as a function of redshift.  As the halo grows in mass
  over time, the virial radius grows, while the core radius shrinks
  due to adiabatic contraction.  The ratio $\rc/r_{200}$ is therefore
  maximized at the time of halo formation, and only diminishes
  thereafter. 
\label{rcore-rvirial}}
\end{figure}

Figure \ref{rcore-rvirial} illustrates the evolution of the halo
structural parameters over time.  Following formation, the halo
continues to accrete matter and steadily grows in mass, at a rate
determined by the initial linear overdensity profile.  By definition,
this growth in $M_{200}$ means that $r_{200}$ grows as well.  Note,
however, that \rc\ decreases as the halo grows, due to adiabatic
contraction of the orbits as the halo potential deepens over time.
Since \rc\ only shrinks in time, while $r_{200}$ grows in time, the
ratio between the two is clearly maximized at the time of formation of
the halo.

\begin{figure*}
\centerline{
\includegraphics[width=0.32\textwidth]{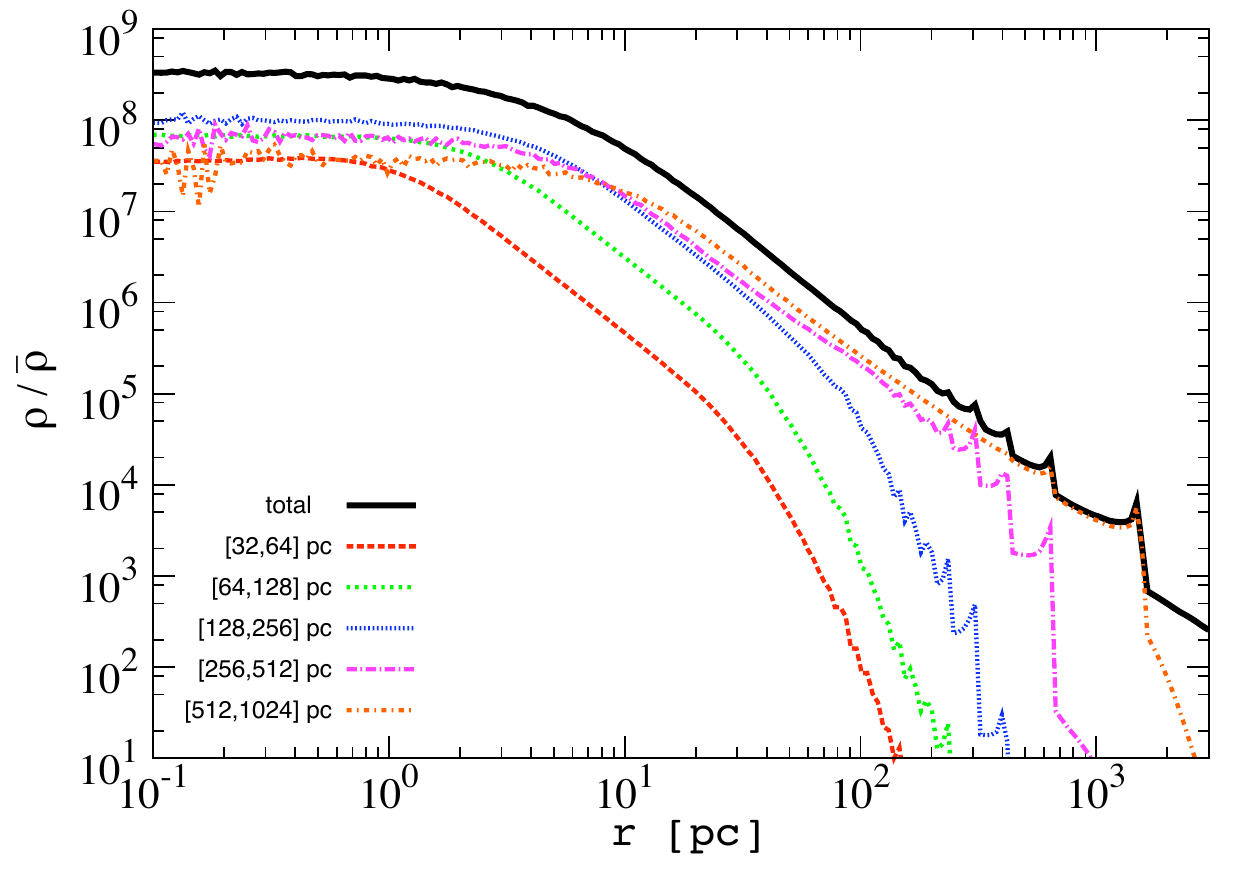}\quad
\includegraphics[width=0.32\textwidth]{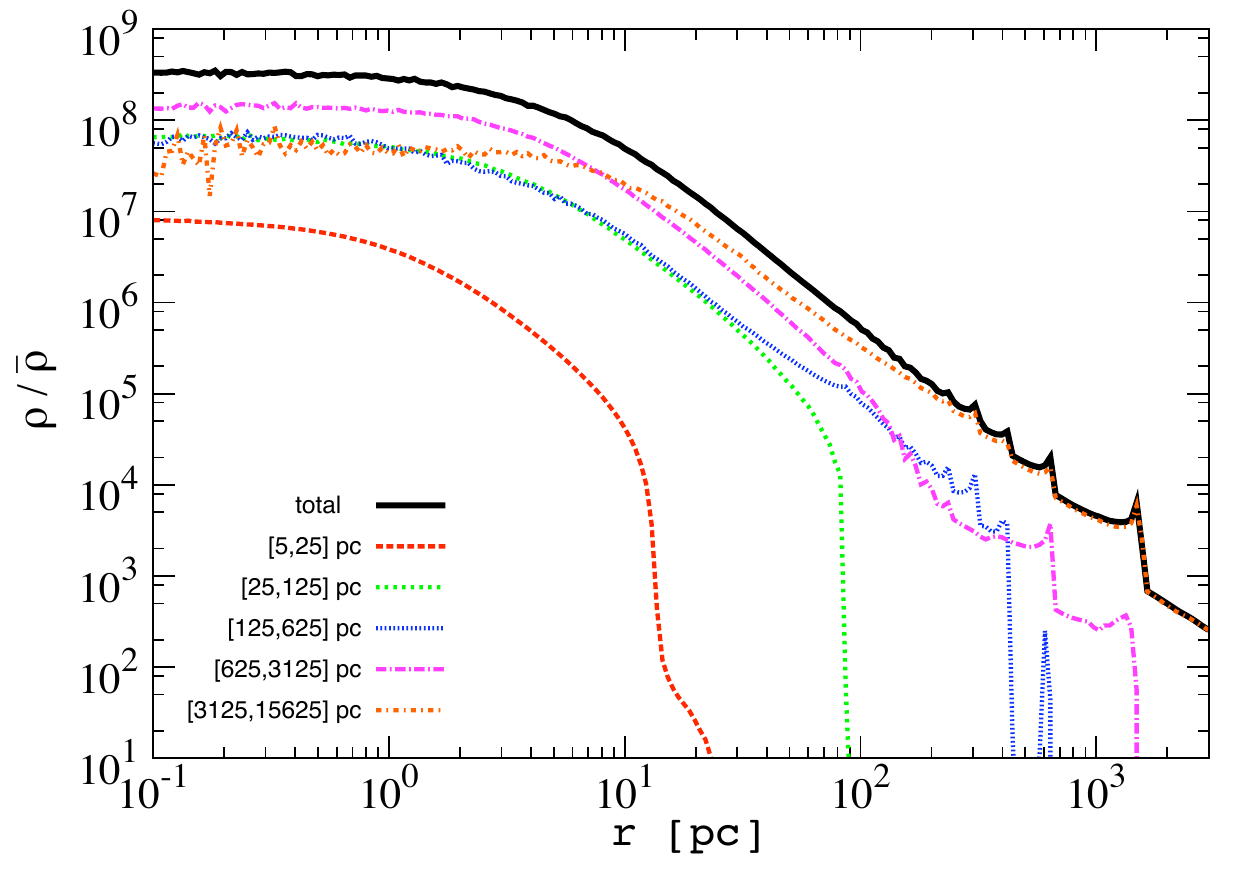}\quad
\includegraphics[width=0.32\textwidth]{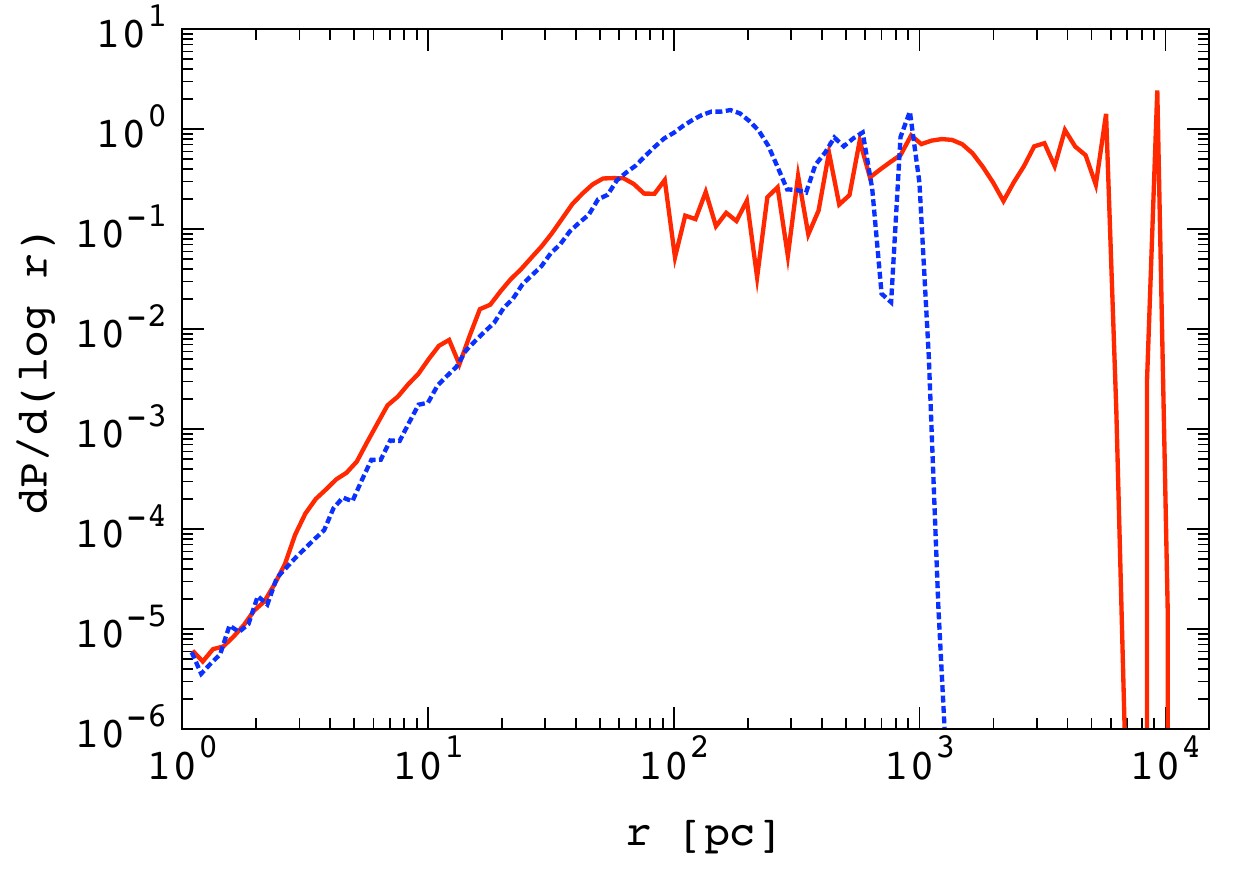}
}
\caption{Breakdown of the mass distribution.  The left panel shows the
  contribution to the mass density $\rho(r)$ at $z=2.6$
  originating from shells at various initial radii at $z=100$.  The
  middle panel shows a similar breakdown, instead binning particles
  based on their radii at turnaround. The right panel shows the
  distribution of initial radii (blue dashed) and turnaround radii
  (red solid) for all particles with $r<\rc$ at $z=2.6$.
  \label{shells}
}
\end{figure*}

Lastly, in Figure \ref{shells} we plot the breakdown of the
contributions to the mass profile from various radii.  The left panel
decomposes the  particles into shells of initial radius at $z=100$,
while the middle panel uses bins of turnaround radius $\rt$. 
At large radii, $r\gg\rc$, the mass is dominated by recently
infalling particles that originated at large Lagrangian radius and
have large turnaround radii, similar to the behavior found in cold
spherical collapse \citep{Fillmore84}.  However, this changes on
scales of order the core radius.  Near \rc, many different shells
spanning decades in radius contribute comparably to the density. When
we bin the particles based on initial, Lagrangian radius, we can see
that each shell has a different core radius, roughly scaling as
$\rc\propto \rL$, as may be expected from the arguments given
in the introduction.  When we bin the particles based on their
turnaround radii, we can see that the typical apoapse for each shell
scales like the turnaround radius.  For shells with $r_{\rm apo}>\rc$,
the enclosed mass profile behaves as 
\begin{equation}
M_{\rm shell}(r) \propto \rL^3 \times \frac{r}{r_{\rm apo}}\times  
\begin{cases} 
1 & r\gg\rc \\ \left(\frac{r}{\rc}\right)^2 & r\ll\rc 
\end{cases}
\end{equation} 
and since $\rc\propto \rL$ for each shell, we see that inside the core
radius, the mass profile deposited by each shell of width $d\log \rL$
scales as $M_{\rm shell}\propto r^3 \times(\rL/r_{\rm apo})\propto 
r^3 \times (\rL/\rt)$.  (For shells with $r_{\rm apo}<\rc$, the
enclosed mass simply behaves as 
$M_{\rm shell}(r) \propto (\rL r/\rc)^3$, of course.)
For cold collapse, there is a
one-to-one relationship between the turnaround radius \rt\ and
the initial radius, that depends on the initial linear density profile
of the peak collapsing to form the halo.  Roughly speaking, if locally
the linear density has slope $\gamma$, in the sense that 
$\delta\propto \rL^{-\gamma}$, then $\rt\propto\rL^{1+\gamma}$ 
\citep{LD10}.  The tight relationship between \rL\ and $\rt$ degrades
somewhat for warm collapse, but we can still use the same basic
scaling.  For shells inside the free-streaming scale, $\rL<\rfs$, the
peak profile is quite flat with local slope $\gamma\approx 0$, and so
$\rt\propto \rL$.  For these shells, $M_{\rm shell}(r)$ becomes
roughly independent of the shell's initial radius for $r<\rc$, as is
seen in Fig.\ \ref{shells}.  For larger radii, $\rL\gtrsim\rfs$, the
slope of the initial profile becomes nonzero, $\gamma>0$, implying
that shells originating from large radius become subdominant inside
the core radius.  Thus, the mass inside the core radius is dominated
by shells with $\rL\lesssim\rfs$ and $r_{\rm apo}>\rc$, and receives
roughly equal contributions per decade within this range, as is seen
in the right panel of Figure \ref{shells}.

\subsection{Dependence on physical parameters}

\begin{figure*}
\centerline{
\includegraphics[width=0.45\textwidth]{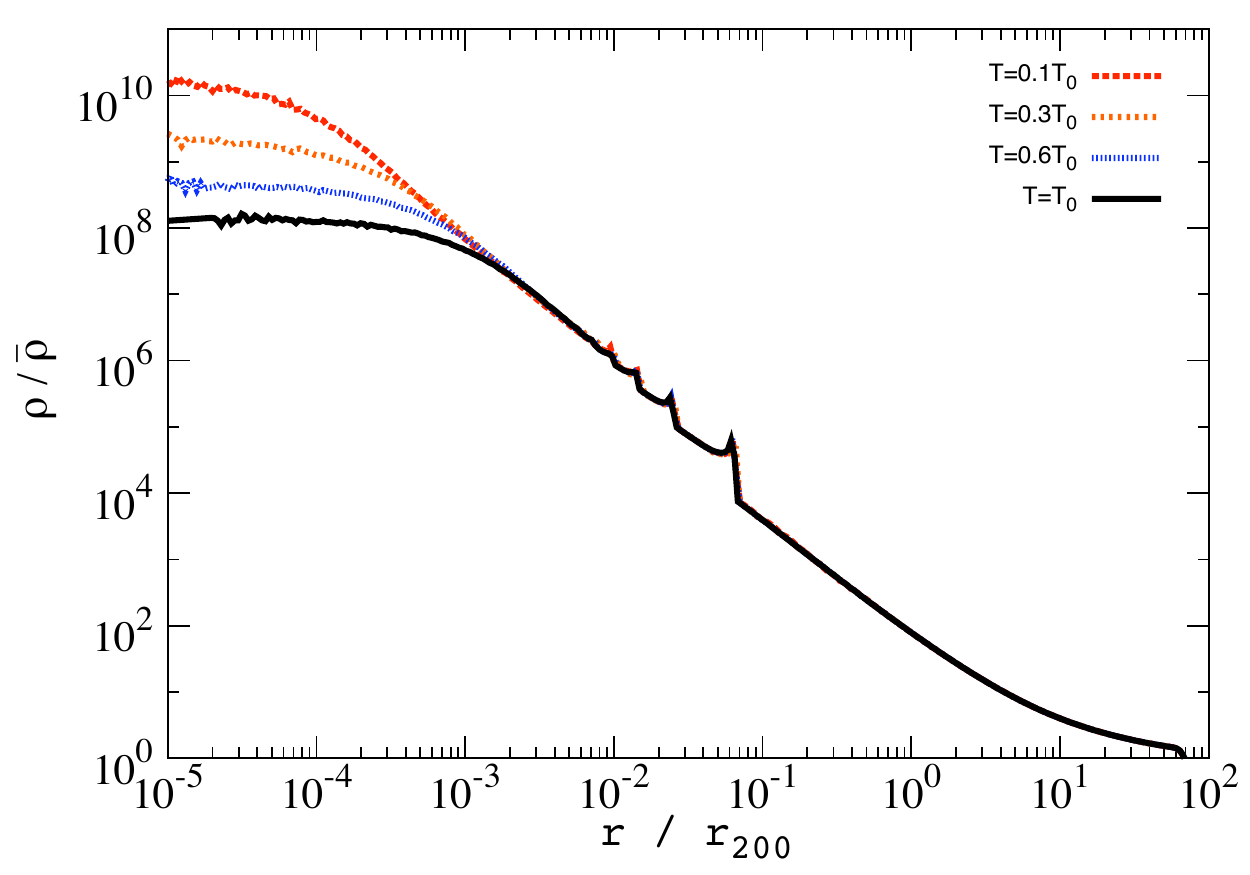}
\quad\includegraphics[width=0.45\textwidth]{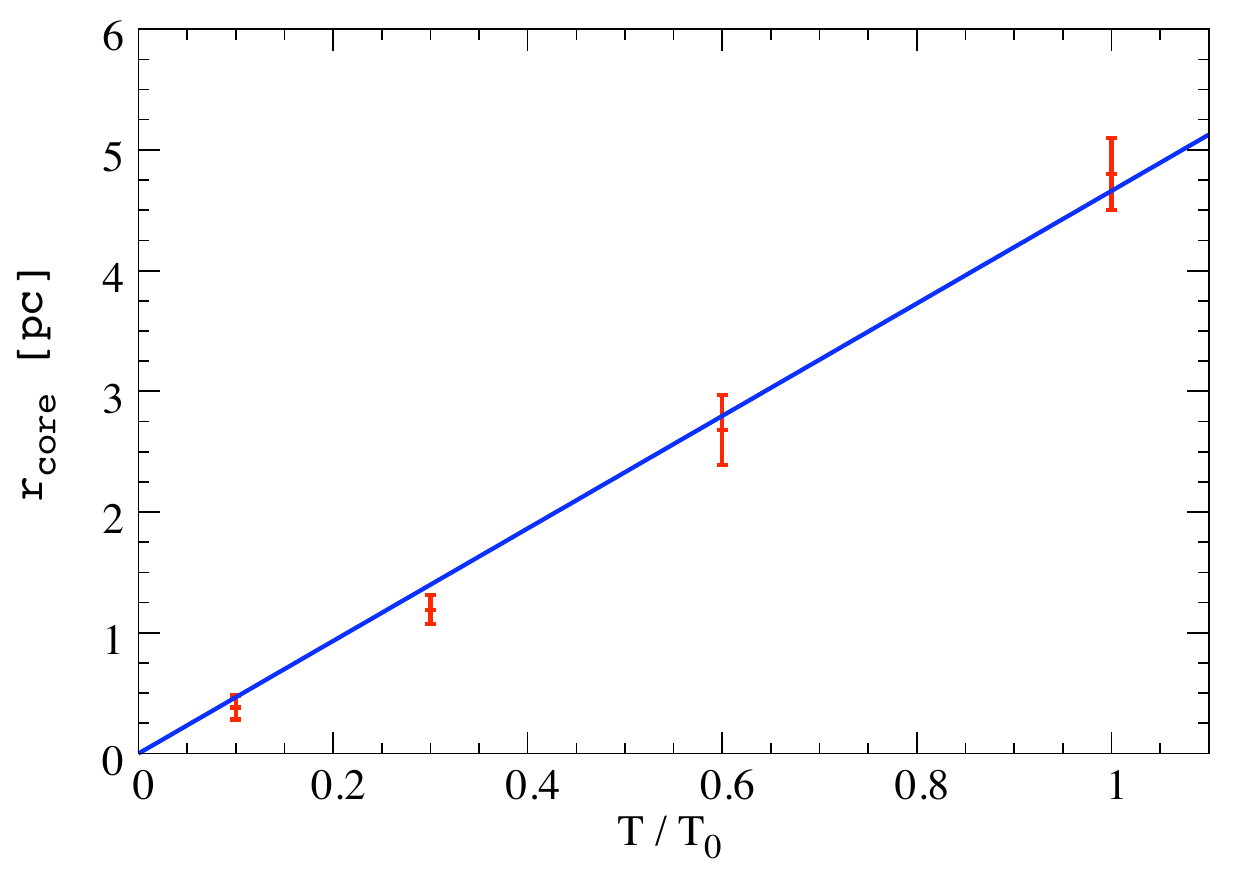}
}
\caption{Temperature dependence of the halo profile for a fixed
  initial peak profile. (Left)
  The black solid curve shows the halo profile at redshift $z=2.7$ for
  $T=T_0$, where $T_0$ is the WDM temperature consistent with the
  linear overdensity of the initial peak.  For comparison,
  the red, orange and blue curves show the profile for $T=0.1 T_0$,
  $T=0.3T_0$ and $T=0.6T_0$ respectively. (Right) The points show the
  fitted core radii for the profiles depicted in the left panel, using
  Eqn.\ (\ref{fit}), while the blue line shows a simple linear
  scaling $\rc \propto T$.
\label{temperature}}
\end{figure*}

Having established the basic features of the halo structure, we now
explore the physics that sets those properties.  The two main
differences between WDM and CDM are the cutoff in the power spectrum,
and the relic thermal velocities of DM particles at the time of halo
formation.  Both of these differences influence the size of halo core
radii.  To disentangle the different effects, we have performed
collapse simulations in which we hold fixed the linear density profile
of the initial peak, but vary the WDM temperature.  This corresponds
to holding fixed the halo assembly history, but varying the random
motions near the time of collapse. The argument given in section 1
(e.g., Eqn.\ (\ref{equa})) would predict that \rc\ would scale
linearly with temperature, and our calculations appear consistent with
this, as shown in Fig.\ \ref{temperature}.  As we vary the
temperature, the overall assembly history and structure of the halo
remains unchanged (e.g., the location and height of the caustics),
however the core radius varies.  We find that a simple linear scaling,
$\rc\propto T$, appears consistent with our simulations.  We note,
however, that this linear behavior breaks down at very high
temperatures, when the particles' random velocities become of order
the Hubble velocity at the time of halo collapse.  In this regime, the
thermal motions are no longer a small perturbation to the particle
dynamics, and the overall collapse of the halo is significantly
modified, unsurprisingly.  Of course, such calculations are not
self-consistent: the large random motions that modify halo collapse at
low redshift would have erased the initial linear density
perturbations responsible for the halo, at a higher redshift.

\begin{figure}
\centerline{\includegraphics[width=0.45\textwidth]{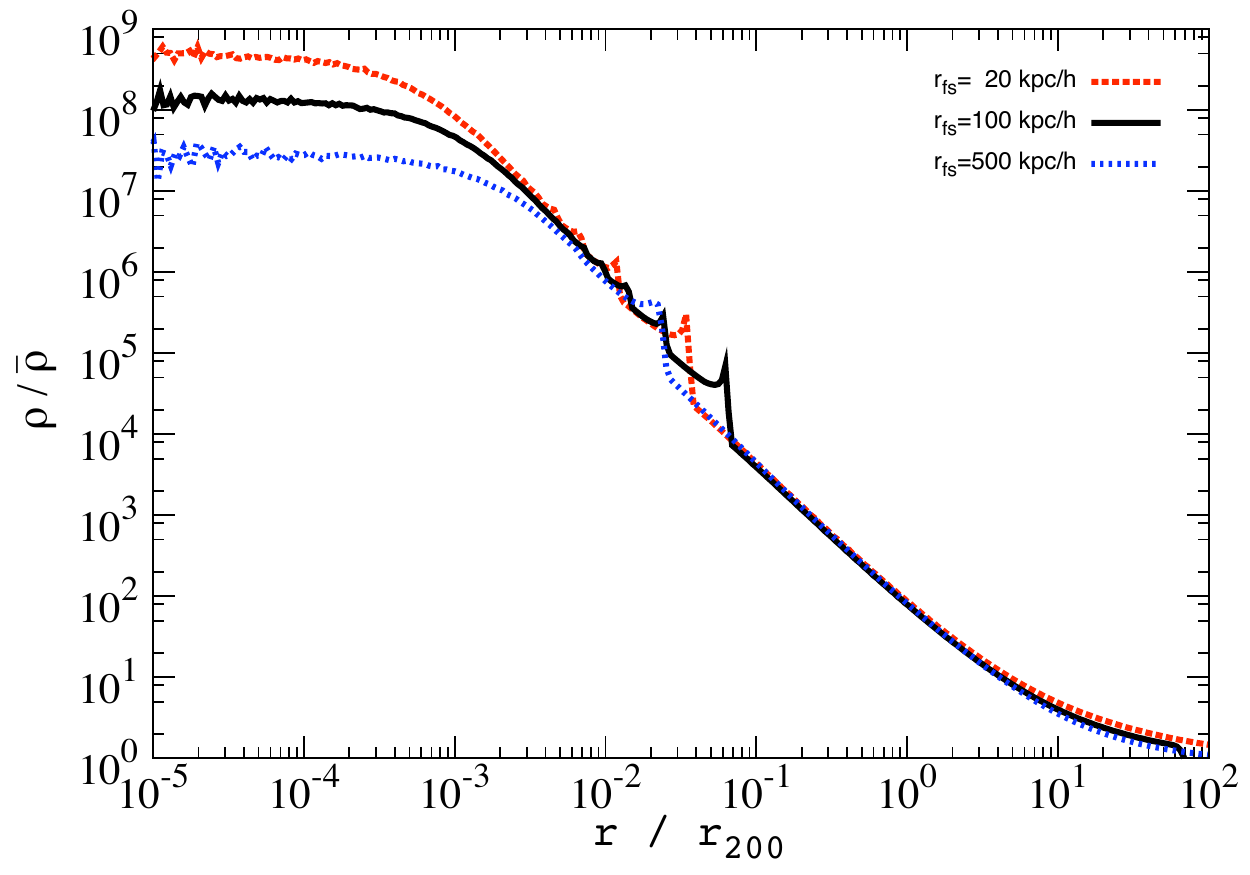}}
\caption{Effect of WDM temperature. Density profiles at $z=2.7$ for
  different initial peak profiles consistent with WDM temperature that
  produce $r_{fs}=$ 20, 100 and 500 kpc/$h$ are plotted in dashed red,
  solid black and dotted blue lines respectively.  
\label{r_fs}}
\end{figure}

This establishes that at fixed initial peak profile (i.e.\ fixed halo
assembly history), the core radius scales linearly with DM
temperature.  It is inconsistent, however, to hold fixed the initial
peak profile while the temperature is varied, since the random thermal
motions of DM particles erase structure at high redshift and modify
the peak profiles in the linear regime of structure formation.
Therefore, we next explore how the core radius behaves as we
self-consistently vary both the WDM temperature and the initial peak
profile.  We know that \rfs\ scales roughly linearly with $T$, and we
have just seen that at fixed \rfs, the core radius \rc\ also scales
linearly with $T$.  Therefore, if \rc\ were independent of the halo
assembly rate, then both \rc\ and \rfs\ would scale linearly with
$T$, and the ratio $\rc/r_{200}$ would be independent of the WDM
temperature, as we argued in Section 1.  Figure \ref{r_fs} shows that
this behavior is not confirmed by our simulations, however.  The
figure shows the results of simulations using temperatures 5 times
larger, and smaller, than our fiducial calculation.  The red solid
curve shows the density profile for $\rfs=20\,h^{-1}\,$kpc, orange
dashed shows our fiducial run with $\rfs=100\,h^{-1}\,$kpc, and blue
dotted shows results for $\rfs=500\,h^{-1}\,$kpc.  In all three cases,
we set the initial peak height so that collapse will occur near
$z=3$.  As expected, $r_{200}$ scales close to linearly with \rfs: the
three simulations give $r_{200}=1.2$, 5.5, and 22.6 kpc at the
formation redshift $z=2.7$.   However, \rc\ does not scale linearly
with $\rfs$: the three simulations give \rc=0.6, 4.8 and 43 pc
respectively.  The ratio $\rc/r_{200}$ is not independent of \rfs, but
instead behaves roughly as $\rfs^{1/2}$ over the range that we have
considered.  Evidently, the core radius depends not only on the DM
temperature at the time of halo formation, but also upon the halo
assembly rate.

\begin{figure}
\centerline{\includegraphics[width=0.45\textwidth]{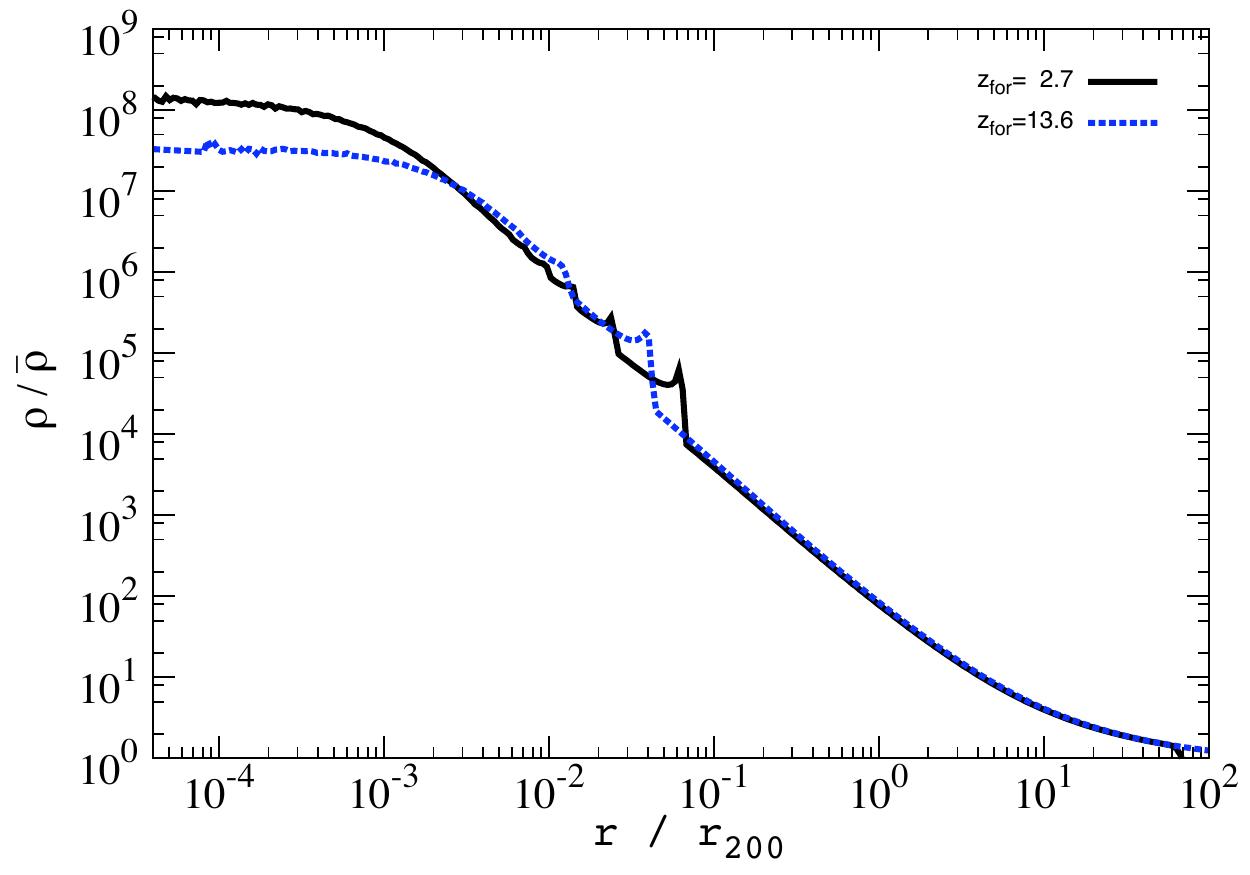}}
\caption{Effect of initial peak height. The solid black line shows the
  profile at collapse time $z=2.7$ for a peak height normalization of
  ${\bar\delta}(\rL=100\,h^{-1}\,{\rm kpc}, z=3)=1.686$. In contrast,
  dashed blue shows the profile at collapse time $z=13.6$ for a peak
  height normalization of ${\bar\delta}(\rL=100\,h^{-1}\,{\rm kpc},
  z=15)=1.686$. 
\label{z_for}}
\end{figure}

Lastly, we examine the dependence of the core radius on the halo
formation time.  We do so, simply by adjusting the height of the
initial peak, holding fixed the WDM temperature and the radial shape
of the peak profile.  Figure \ref{z_for} shows one example, comparing
our fiducial simulation (with $z_{\rm form}=2.7$) with a run using the
same WDM temperature, but with initial peak height a factor of 4
larger.  The later simulation has $z_{\rm form}=13.6$.  The halo masses of
the two simulations are similar, as expected: $r_{200}=1.3\,$kpc for
the $z=13.6$ halo, compared to $r_{200}=5.5\,$kpc for the fiducial
$z=2.7$ halo.  The core radius for the earlier-forming halo is \rc=2.8
pc, compared to 4.8 pc for the fiducial halo, so that the ratio
$\rc/r_{200}$ changes by a factor of 2.8.  The simple argument given
in section 1 would have predicted that $\rc/r_{200}$ would scale as
$(1+z_{\rm form})^{1/2}$, whereas our simulation appears more
consistent with a scaling $\rc/r_{200}\propto(1+z_{\rm form})^{2/3}$.
This is only based on one comparison, of course.  This is the result at
the formation time; at $z=0$ the ratio $\rc/r_{200}$ would be smaller
by at least a factor of $1/(1+z_{\rm form})$, as we argued earlier.

\section{Discussion and summary}

We have investigated the formation of halos in warm dark matter
cosmologies.  Our study of spherical collapse of WDM halos indicates
that core radii do indeed arise in these cosmologies, as expected from
simple phase-space arguments \citep{Tremaine79}.  However, we find
that WDM core radii are generically small, typically of order
$10^{-3}$ of the halo virial radius at the time of formation, and
considerably smaller following formation.  This is for halos forming
at the cutoff scale; higher mass halos will have substantially smaller
$\rc/r_{200}$.  We have investigated the dependence of the core radius
on various physical parameters such as the WDM temperature, halo
formation redshift, and halo mass.  For the allowed range of WDM
temperatures (e.g. $\Mfs \lesssim 10^9 M_\odot$), the core radii of
halos observed at $z=0$ are generically expected to be far smaller
than the core sizes measured in certain LSB galaxies, with
$\rc/r_{200}\approx 0.05$.  

Our calculations have all assumed spherical symmetry, whereas halo
formation in both CDM and WDM cosmologies is highly nonspherical.  We
would argue, however, that our conclusions regarding WDM cores are
likely to be valid for non-spherical collapse as well.  One line of
evidence supporting this argument is the fact that the
pseudo-phase-space density profiles of our halos are quite similar to
the profiles of halos in CDM simulations, with $\rho/\sigma^3\propto
r^{-2}$, roughly speaking \citep{Ludlow10}.  This similarity is
presumably a consequence of the virial theorem, which ensures that
$\rho/\sigma^2\sim r^{-2}$.  For WDM halos, the power-law rise of
the pseudo-phase-space density towards small $r$ saturates when
$\rho/\sigma^3$ 
approaches the Tremaine-Gunn bound, and this saturation will occur for
both spherical and nonspherical collapse.  Now, we would expect the
value of the pseudo-phase-space density to be similar at the virial
radius for both spherical and non-spherical collapse, since the halo
mass and virial radius are by definition the same in the two cases,
and so $\rho\sim M/4\pi r^3$ and $\sigma\sim (GM_{200}/r_{200})^{1/2}$
will be similar for the two cases.  We have noted that $\rho/\sigma^3$
rises as roughly $r^{-2}$ inside the virial radius in spherical and
nonspherical collapse, and in both cases the core radius occurs where
$\rho/\sigma^3$ approaches the Tremaine-Gunn bound.  So we have good
reason to believe that halo core radii will not be significantly
larger for nonspherical collapse than for spherical collapse, just
because the $\rho/\sigma^3$ profiles appear similar.

The other possible loophole in our argument is that we have assumed
that no halos form below the cutoff scale in the power spectrum.
N-body simulations have not yet conclusively determined whether or not
halos with $M\ll\Mfs$ arise in WDM cosmologies, due to numerical
difficulties associated with simulating truncated power spectra
\citep{Wang07}.  We have begun investigating this issue, and our
preliminary results indicate that halos may form below the cutoff
scale, though it is unclear whether they can form in sufficient
numbers to account for observed LSB galaxies.

Our results indicate that warm dark matter cosmologies cannot produce
halos with core radii large enough to account for the density profiles
of observed LSB galaxies.  This would suggest that the origin of these
observed cores lies within astrophysics, rather than particle physics.

\acknowledgments{We thank Manoj Kaplinghat, Mark Vogelsberger and
  Larry Widrow for useful discussions.  FVN thanks CITA and the CfA
  for hospitality during the course of this work. ND is supported by
  CITA, and FVN is supported by CSIC pre-doctoral grant JAE 2008.  
}


\end{document}